\documentclass[twocolumn,superscriptaddress,showpacs,floatfix]{revtex4}
\usepackage{graphicx}
\usepackage{epstopdf}
 \begin{document}
\title{Strong  associations between  microbe phenotypes and their network architecture}
\author{Soumen Roy}
\affiliation{Dept of Medicine and Institute for Genomics and Systems Biology, The University of Chicago, Chicago, IL 60637}
\affiliation{Dept of Mechanical and Aerospace Engineering, University of California,  Davis CA 95616}
\email{soumen@uchicago.edu} 
\author{Vladimir Filkov}
\affiliation{Dept of Computer Science,  University of California,  Davis CA 95616}
\email{filkov@cs.ucdavis.edu}
\pacs{87.18.Vf, 89.75.Hc, 07.05.Rm}
\begin{abstract}
Understanding the dependence and interplay between architecture and function in  biological networks  has great relevance to disease progression,  biological fabrication  and biological systems in general. We propose methods to assess the association of various microbe characteristics and phenotypes with the topology of their networks. We adopt an automated approach to characterize metabolic networks of $32$  microbial species using $11$  topological metrics from  complex networks. Clustering allows us to extract  the indispensable,  independent  and informative metrics. Using hierarchical linear  modeling, we identify relevant subgroups of these metrics and establish that they associate with microbial phenotypes surprisingly well.  This work can serve as a stepping stone to cataloging biologically relevant topological properties of networks and towards better modeling of phenotypes. The methods we use can also be applied to networks from other disciplines. 
\end{abstract}
\maketitle
A prime goal of systems biology is to discover emergent properties that may be unraveled when  a systemic view  is adopted to gain a comprehensive understanding of many processes that occur in biological systems. The reductionist approach which has held sway in biology over the past several decades has successfully identified the key components in living systems and many interactions among them. However,  it almost never presents a holistic understanding of how the systemic properties emerge. It is now becoming increasingly clear that the functioning of biological systems  depends crucially on their complex underlying structure~\cite{PaolaOliveri04222008}. This complexity is the consequence of  numerous interconnected,  dynamic and nonlinear interactions among the plethora of elements,  like genes,  proteins,  and metabolites.   But the importance of biological networks lies beyond their being the most visible signatures of complexity.  Understanding the dependence and interplay between architecture and function in biological networks  has great relevance to disease progression,  bio-fabrication  and biological systems in general.

The central issue, then, is to discover whether networks encode systemic events and the precise manner in which they do so.  Ideally,  we would like to understand and modify the complex behavior of biological networks,  which is contingent on the proper level of modeling of their molecular interactions. To model the systemic or emergent properties, one would have to involve critically, the interdependencies among interactions and other organizational patterns, on a local level (e.g. network motifs) as well as global level (e.g. modularity). Recent research in complex  systems and networks has  presented opportunities to properly mine and thence exploit the architectural  interdependence in networks~\cite{sharan,russel,koide}. 

Multiple metrics exist in complex networks and various studies have utilized one or few of them at a time,  to  characterize biological networks.  Significant research has been done to examine various  topological properties of different networks using computational and analytical methods. It has been found that many biological networks (just like other empirical networks) may have power-law degree distributions~\cite{barabasi: cellular},  are modular~\cite{modularity} and hierarchical~\cite{hierarchy},  and have specific distributions of  topological features which can be used to characterize them ~\cite{motif1, motif2, erg}. In addition, topological properties have been used to predict missing edges in networks ~\cite{clauset} and viability of mutant strains~\cite{wunderlich}. 

In this rapid communication,  we show that various topological metrics (which are the signature of complex network architecture), associate with microbe characteristics and phenotypes to a surprisingly high degree. We undertake an automated approach using various topological metrics from complex networks to characterize a collection of various kinds of biological networks and show how these metrics associate strongly with microbe characteristics. Specifically, 
(i) using publicly available data we collect and cross-reference metabolic networks for $32$  different  microbes via ten different quantifiable characteristics and phenotypes, 
(ii) we use a suite of  $11$   complex network metrics,  so as to comprehensively compare all  $32$   networks simultaneously,  allowing for a much more in-depth evaluation of network models~\cite{graphlets} than is possible with the usually existing practice of comparing one or two particular properties,  most commonly the degree distribution,  
(iii) we show that most of the network metrics we use are independent and that multiple metrics are necessary to characterize the variability in networks meaningfully, 
(iv) via a hierarchical linear modeling approach, we identify subsets of network parameters which associate strongly with various  microbe  characteristics and phenotypes. 
By presenting these strong associations and exhibiting the necessity of multiple metrics to do so, this work is a step forward toward a systemic cataloging of the methods and properties of biological networks that are relevant to the underlying biology, and towards better modeling of emergent biological properties. 

The  microbe characteristics or phenotypes that are explored in this work are:   (1)  microbe class (MC), (2) genome size (GS), (3) GC content (GC), (4) modularity $(Q)$,  (5)  number of such modules $(N_Q)$,  (6)  motility $(MO)$,  (7) competence$(CO)$  and  whether these microbes are (8) animal pathogens ($AP$),  (9) strict anaerobes $(AN)$, or, (10) extremophiles $(EX)$. 

Microbes are normally classified as archaea or bacteria ~\cite{woese}. 
Genome size alludes to the sum total of DNA contained within one copy of a genome. The usual measure of it is in terms of mass in picograms or the total number of nucleotide base pairs (commonly as millions of base pairs,  or megabases). Intriguingly,  an organism's genome size is not directly proportional to its complexity and a few microbes have much more DNA compared to other microbes. In this context, it is interesting to point out that the association between genome size and topological metrics of the networks are among the strongest of all phenotypes explored in this work. GC content is the percentage of nitrogenous bases on a DNA molecule which is either cytosine or guanine (and not thymine or adenine).  Data for genome size and GC content was obtained from the NCBI Entrez genome project database~\cite{gc}. With regard to  biological networks, modularity  is defined as the fraction of edges within modules less the expected fraction of such edges. We use a recent algorithm~\cite{leicht}  in determining the community-structure in networks which incorporates edge directionality. 
Until recently, the most common approach to modularity in complex networks literature has been to simply ignore edge direction and apply methods developed for community discovery in undirected networks. However,  this discards potentially useful information contained in edge directions, which is most commonly a very biologically relevant criterion. It should be noted that modularity is 
intimately connected to function in biology as the modules typically correspond to genetic circuits or pathways~\cite{modularity,  mod2}.  Therefore, we include it here as a phenotypic property rather than as a variable. In  scenarios where modularity lacks apparent connection to function, it is more appropriate to treat $Q$ and $N_Q$ as input variables.

Motility allows microbes to move toward desirable environments and away from undesirable ones. Competence denotes the ability of a cell to take up extracellular DNA from its environment. Anaerobic organisms are those that do not require oxygen for growth and may even die in its presence.  Extremophiles are organisms which thrive in or require extreme physical or geochemical conditions, in which majority of life on Earth cannot survive. Data for phenotypes (6) to (10) have been  compiled from Ref.~\cite{arkin}.  While $GS, GC, Q, N_Q$ can take on any value, the rest of the microbe characteristics or phenotypes are ``binary" (e.g., a microbe is either an archaea or a bacteria; either aerobic or anerobic, etc.).   

We used  metabolic networks of  $32$   different  microbes based on data deposited in the WIT database~\cite{wit43}. This database contains metabolic pathways that were predicted using the sequenced genomes of several  organisms.  The nodes in these networks are enzymes,  substrates and intermediate complexes,  while  edges represent sequences of reactions in the organism's cells. (We had to exclude the following three microbial species: {\it A. actinomyc.,  R. caps.} and {\it M. thermoautot.},  from the original collection because many of the  microbe characteristics or phenotypic data do not seem to be publicly available for them.) The network sizes vary from $595$ nodes and $1354$ edges to $2982$ nodes and $7300$ edges. 

We calculated a suite of  $11$    important complex network attributes across  all $32$    networks. These are:   the number of {\em nodes}, $N$, and {\em edges}  in the network and the first three standardized moments (mean,  standard deviation,  and skewness) of the distributions of   geodesic~\cite{dijkstra},  betweenness coefficient~\cite{betweenness}, and,  degree, of the network;  respectively denoted as $ geo_1,  geo_2,  geo_3,  betw_1,  betw_2,  betw_3$,  $deg_1,  deg_2,  deg_3$. The importance of studying the higher moments of distributions is well known in physics~\cite{georges}.   The {\em Geodesic} was calculated by using the Dijkstra Algorithm~\cite{dijkstra}. 
For normalization, we subtract the mean value of a metric (over all species) and then divide by the standard deviation of the metric (over all species), for all networks. Some of our metrics are robust to measurement errors.  Observing the system (i.e. network) from multiple angles,  provides a measure of robustness against noise (false positives and false negatives).

\begin{figure}
\label{figure: heatmap}
\includegraphics[width=2.5in, angle=-90]{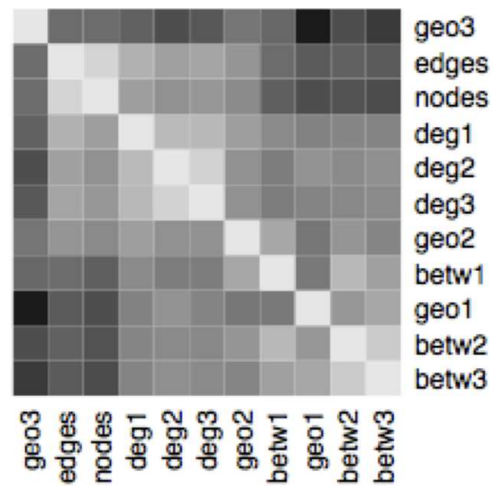}
\label{fig:heatmap}
\caption{The heatmap over network metrics}
\end{figure}

\begin{table*}[htbp]
\label{tablesignif}
\begin{tabular}{|c|c|c|c|c|c|}
\hline
&Range (min,max)&{${\rho}^{}_{best}$} & {\em $\langle{\rho}^{}_{random}\rangle$} & p-value & Best Model Variables\\
\hline
MC & Binary &0.113 & 0.507 & $< 3\times10^{-5}$ &$N, edges, geo_1, geo_2, geo_3, betw_1, betw_2, betw_3, deg_1$\\
 
GS & (0.58, 6.3) &0.476 &  1.302 & $< 10^{-6}$  & $N, edges, betw_1,  betw_2, betw_3,  deg_2, deg_3$\\

GC & (28.2, 66.6) &0.763 &  1.158 &  $<9.8\times10^{-5}$ & $N, edges, geo_1, geo_2, geo_3, betw_1$\\

Q & (0.59, 0.69) & 0.005 &  0.033 &  $< 10^{-6}$ & $N, edges, geo_2, geo_3, betw_1, betw_3, deg_1, deg_2$\\

$N_Q$ &(14, 35)  &2.102 &  6.413 &  $< 10^{-6}$ &  $N, edges, geo_1, geo_2, geo_3, betw_1, deg_1, deg_2$\\

MO & Binary &0.315 &  0.577 &  $<1.4\times10^{-5}$ & $N, edges, betw_3, deg_1, deg_2, deg_3$\\

CO& Binary &0.158 & 0.683 &  $< 9\times10^{-6}$ & $N, edges, geo_1, geo_2, geo_3, betw_1, betw_3, deg_1, deg_2, deg_3$\\

AP & Binary &0.325 &  0.567 &  $< 10^{-6}$ & $geo_1,  geo_2, betw_3,  deg_2, deg_3$\\

AN & Binary &0.359 &  0.495 &  $<2.66\times10^{-4}$ & $edges, geo_1, geo_3, betw_1, betw_2, betw_3, deg_3$\\

EX & Binary &0.284 &  0.540 &  $< 10^{-6}$ & $geo_1, geo_2, betw_3, deg_1, deg_2, deg_3$\\
\hline
\end{tabular}
\caption{Exploring the association of microbe characteristics and phenotypes with network metrics:  Microbe class (MC);  Genome size (GS);  GC content (GC); Modularity $(Q)$;  Number of modules$(N_Q)$; Motility (MO); Competence (CO); and whether the microbes are Animal Pathogens (AP), Strict Anaerobes (AN) or Extremophiles (EX).}
\end{table*}

The degree of overlap,  or dependence,  between the attributes when characterizing networks can be accurately assessed by a symmetric heatmap,  showing the pairwise correlations  of the network metrics over all the networks. Fig.~\ref{fig:heatmap}1 shows the heatmap over network attributes. We start with a $32$  dimensional vector (which is the number of microbes studied) for each of the $11$  metrics. Thus, we have $11$  points in the $32$  dimensional vector space. We then calculate the correlation between all pairs of these $11$  points, and color-code the distance. White indicates perfect correlation while black indicates anti-correlation.  
The rows (and by symmetry the columns) are arranged automatically so that the rows most similar are placed next to each other, as determined by the hierarchical clustering algorithm implemented in the {\em heatmap} package of the R system~\cite{R-project} (as any other clustering scheme, this one too has its limitations, e.g. in the placement of the edges and nodes columns, which could arguably  be swapped). Thus, the map allows us to identify clusters of ``similar'' network attributes by looking for blocks of light-colored squares along the diagonal of the figure.  Since there is only a small amount of clustering along the diagonal,  it follows that the network attributes we have chosen are relatively independent,  and thus,  they all provide information to our analysis. 

To find how well the organism phenotype associates with the underlying network architecture, we consider our $11$  network metrics (which can be regarded as characteristics of the architecture)  and model each phenotype as a linear combination of these metrics. It should be especially noted that the basis of linear modeling is not to imply that the dependent variables are the cause and the explanatory variables are the effect,  but that there is a significant association between these variables. 
Anticipating that not all metrics will be pertinent to each phenotype, and, in general, to avoid over-fitting we use {\em hierarchical linear regression} methods to model the phenotypes as linear combinations of subsets of the network metrics. To identify the best model we start by assuming a linear dependence on all  $11$  variables, because we do not  know initially which ones associate better than others. We then iteratively proceed to exclude variables whose absence  improves or  does  not significantly alter the quality of the resulting model (we used a specific implementation of this iterative procedure through the {\em step()} function of the R system~\cite{R-project}).  The model selection is guided by minimizing the well-known Akaike Information Criterion~\cite{AIC}   denoted here as   $\alpha$,  a standard measure in statistics allowing for selection among various nested models. $\alpha$  scores a model based on its goodness-of-fit to the data and penalizes models having many parameters. If $k$ is the number of parameters in the statistical model, and $L$ is the maximum log-likelihood for the estimated model, $\alpha$ is defined as:
\begin{equation}
\alpha = 2k - 2\ln(L)\
\end{equation}

Thus, we identify the smallest number of independent and indispensable network metrics that can be associated with the microbe characteristic or phenotype. The results for the best model for each phenotype are given in Table 1\ref{tablesignif}. We use the Root Mean Square Error, $\rho$,  which is a measure of the goodness-of-fit of our model associations and the experimental data. $\rho$ of an estimator $\hat{X}$ with respect to the estimated parameter $X$ is defined as the square root of the mean squared error, 
\begin{equation} 
{\Large{\rho}} (\hat{X}) = 
{\sqrt{{E}((\hat{X}-X)^2)}} .  
\end{equation}
We also report the significance of the best model, which we obtain by the linear hierarchical modeling procedure discussed above by bootstrapping  with respect to the same model and using a random permutation of the observed data. We measure the $\rho$ of these random models, $\rho_{random}$, and how many times (or whether at all) $\rho_{random}  < \rho_{best}$, where $\rho_{best}$ is obviously the $\rho$ of the best model. The number of times this happens is reflected in the normalized  significance. We observe $10^6$ such random permutations, for each microbe phenotype.  We also performed an analysis of variance (ANOVA) of the difference of our model with fewest dependent variables versus the model with all  $11$  variables, and the difference was not significant. 

For half of the microbe phenotypes in this study (GS, $Q$, $N_Q$, AP  and EX), we do not come across a single instance where the $\rho_{random}$  is less than $\rho_{best}$ for that phenotype.  For each of these five phenotypes and also for the rest of the ones considered in this study,  $\rho_{best}$ is always less than $\langle \rho_{random} \rangle$, with very low p-values.  This, thus indicates a strong association of organism phenotypes with the relevant network metrics, in general.   

There are some other facts which are observable from Table 1\ref{tablesignif}: (i) there is no supremely important single metric associated with each and every phenotype studied here, and, (ii)  in the present study, this of course rules out one or more set of metrics associated  with more than one phenotype(s). Albeit, the latter occurrence does not automatically follow from the former if one or more metrics is consistently observed to be associated with all phenotypes.  These facts, however, attest to the indispensability of the simultaneous study of multiple network metrics.  It is notable that the association patterns are non-trivial, even when the microbe phenotype or characteristic is simply binary, as opposed to the case, when it possesses a range of values. The dependence of the prediction quality on the number of metrics is also not readily ascertainable. For example in AP,  five of the $11$  metrics seem to be needed for sufficient representation, while eight are required for $Q$ and $N_Q$. However, with six metrics for GC and MO, or nine for MC the prediction quality is apparently not enhanced.

Interestingly,  the orthogonality of  the geodesic and betweenness metrics which we established before is reflected by their consistent appearance in the association results.  It is entirely possible that the association of other network metrics, which are not a part of this study, with these or other phenotypes or organism characteristics could be particularly strong.  Exhaustive studies with the inclusion of such metrics should bear out this fact. Here we focused on metrics that have been shown to be biologically pertinent. The approaches adopted here are scalable and can easily accommodate other important metrics, which could be unraveled in future as a result of the continuous ongoing research in network theory.

The importance of this study is in justifying that suitably identified groups of network metrics, can and should be used to meaningfully model and study organism characteristics. Most immediately, the results can be used to build more sophisticated and even predictive models of organism phenotypes, based on their network architecture. These results are also a good starting point for classification or cataloging of biologically relevant topological features, that can eventually yield vocabularies which cross-reference  topology with biological function. While still far away,  we expect such tabulated and  well-described architectural features to be akin to biological markers in other empirical data. In this sense, our work is a modest step toward understanding the precise nature of interdependence between function and topology in biological networks. Followup modeling and simulations could give valuable insight into a wide range of far-reaching issues like the effect of topology on the design and  evolution of networks. The comprehensive ``look-up scheme",  elucidated with  the present set of biological networks,  could also be helpful for other real-world complex networks in general. Of course,  the measures need not be the same as those above and will depend on the nature and topology of the network. 

It is well known that various centrality measures play an important role in networks and in some cases (e.g.,  in the global airline network~\cite{GuimPNAS05}),  few nodes which have relatively low degree but high betweenness could be very special. Nodes with high betweenness can act as bottlenecks for information passage and the role of betweenness is well known in epidemiology, information and wireless or sensor networks. The role of betweenness in biological networks is being thoroughly exploited in recent times~\cite{kevin-science}. However,  to our knowledge, there is  almost no in-depth work in literature, investigating  the role of higher moments of the betweenness distribution in biological networks.  The present work underlines the importance of such studies.

We are grateful to Z. Saul for help in data gathering for some of our networks. We thank E. Leicht and M. Newman for providing us the codes on modularity structure in directed networks.  This work was funded in part by the NSF under Grant No. IIS-0613949.


\begin{thebibliography}{10}

\bibitem{PaolaOliveri04222008}
P Oliveri, Q Tu, EH Davidson,
Proc Natl Acad Sci USA, {\bf 105}, 5955 (2008)

\bibitem{sharan}R Sharan and T Ideker, Nat. Biotech., {\bf 24}, 427 (2006)
\bibitem{russel}RB Russel and P Aloy, Nat. Chem. Biol., {\bf 4}, 666 (2008) 
\bibitem{koide}T Koide, WL Pang  and  NS Baliga, Nat. Rev. Microbiology, {\bf 7}, 297 (2009)

\bibitem{barabasi: cellular}
H Jeong {\it et al.},  
Nature,  {\bf 407},   650  (2000).

\bibitem{modularity} N Kashtan and U Alon,  
Proc Natl Acad Sci U S A.  {\bf 102},  13773 (2005)

\bibitem{hierarchy}
M Sales-Pardo, R. Guimer{\`a}, AA  Moreira,  and LAN Amaral
Proc Natl Acad Sci U S A.,  {\bf 104} 15224 (2007)

\bibitem{motif1} R Milo  {\it et al.},  
Science,  {\bf 303},  1538 (2004)

\bibitem{motif2} R Milo  {\it et al.},  
Science,  {\bf 298},   824 (2002)

\bibitem{erg} ZM Saul and V Filkov,  
Bioinformatics,   {\bf 23},  2604 (2007)

\bibitem{clauset} A Clauset, C Moore and MEJ Newman, Nature, {\bf 453}, 98 (2008)

\bibitem{wunderlich} Z Wunderlich, and LA Mirny, 
Biophys J., {\bf 15}, 2304 (2006)

\bibitem{graphlets}
V Filkov, ZM Saul, S Roy, RM D'Souza and PT Devanbu, 
Europhysics Letters, {\bf 86},  28003 (2009)

\bibitem{woese}
 CR Woese and GE Fox, PNAS,  {\bf 74},  5088 (1977)

\bibitem{gc} \url{http: //www.ncbi.nlm.nih.gov/genomes/lproks.cgi}

\bibitem{mod2}
G Schlosser and G Wagner,  
\newblock Modularity in Development and Evolution. 
The Univ. Chicago Press,  Chicago  (2004)

\bibitem{leicht}
EA Leicht and MEJ Newman, 
Phys. Rev. Lett,  {\bf 100},  118703 (2008)

\bibitem{arkin}
AH Singh, DM Woolf, P Wang and AP Arkin, 
Proc. Natl. Acad. Sci. USA, {\bf 105}, 7501 (2008)

\bibitem{wit43} R Overbeek  {\it et al.},  
Nucleic Acids Res. {\bf 28},  123 (2000)

\bibitem{dijkstra} EW Dijkstra,  
Numerische Mathematik,  {\bf 1} (1959),  S 269.

\bibitem{betweenness}
LC Freeman,  
Sociometry,  {\bf 40},  35 (1977)

\bibitem{georges} JP Bouchaud and A Georges,  
Physics Reports {\bf 195}  127 (1990)

\bibitem{R-project} \url{http: //www.r-project.org/}


\bibitem{AIC} H Akaike, 
IEEE Trans. on Automatic Control,  {\bf 19},  716 (1974)


\bibitem{GuimPNAS05}
R Guimer{\`a}, S Mossa, A  Turtschi, LAN Amaral,  
Proc. Natl. Acad. Sci. USA,  {\bf 102},  7794 (2005)

\bibitem{kevin-science}
J Liu {\it et al.}, Science, {\bf 323}, 1218 (2009)  



\end{thebibliography}
\end{document}